\documentclass[12pt,preprint]{aastex}
\usepackage{graphics}
\input epsf
\begin{document}
\title{Star Formation from Galaxies to Globules}
\author{Bruce G. Elmegreen
  \affil{IBM Research Division, T.J. Watson Research Center,
    P.O. Box 218, Yorktown Heights, NY 10598, USA}
}

\begin{abstract}
The origin of the empirical laws of galactic-scale star formation
are considered in view of the self-similar nature of interstellar gas
and the observation that most local clusters are triggered by specific
high-pressure events.  The empirical laws suggest that galactic-scale
gravity is involved in the first stages of star formation, but they do
not identify the actual triggering mechanisms for clusters in the final
stages. Many triggering processes satisfy the empirical laws, including
turbulence compression and expanding shell collapse.  The self-similar
nature of the gas and associated young stars suggests that turbulence is
more directly involved, but the energy source for
this turbulence is not clear and the small scale morphology of
gas around most embedded clusters does not look like a random turbulent flow.
Most clusters look triggered by other nearby stars.  Such a prominent
local influence makes it difficult to understand the universality of
the Kennicutt and Schmidt laws on galactic scales.

A unified view of multi-scale star formation avoids most of these
problems.  The Toomre and Kennicutt surface density
thresholds, along with the large-scale gas and star formation morphology,
imply that ambient self-gravity produces spiral arms and giant
cloud complexes and at the same time drives much of the turbulence
that leads to self-similar structures.
Localized energy input from existing clusters and field supernova drives
turbulence and cloud formation too, while triggering clusters directly
in pre-existing clouds.  The hierarchical structure in the gas made by
turbulence ensures that the triggering time scales with size, thereby
giving the Schmidt law over a wide range of scales and the size-duration
correlation for young star fields.

Reanalysis of the Schmidt law from a local point of view suggests that
the efficiency of star formation is determined
by the fraction of the gas above a critical density of around $10^5$
m(H$_2$) cm$^{-3}$.  Such high densities probably result from turbulence
compression in a self-gravitating gas, in which case their mass fraction
can be estimated from the density distribution function that results
from turbulence.  For Wada \& Norman's log-normal function that arises
in whole-galaxy simulations, the theoretically predicted mass fraction
of star-forming material is the same as that observed directly from the
galactic Schmidt law, and is $\sim10^{-4}$.  
 
The unified view explains how independent star formation processes can
combine into the empirical laws while preserving the fractal nature
of interstellar gas and the pressurized, wind-swept appearance of most
small-scale clouds.  Likely variations in the relative roles of these
processes from region to region should not affect the large-scale average
star formation rate.  Self-regulation by spiral instabilities and star
formation ensures that most regions are in a marginally stable state
in which turbulence limits the mass available for star formation and
the overall rate is independent of the nature of the energy sources.
In this sense, star formation is saturated to its largest possible value
given the fractal nature of the interstellar medium.

\end{abstract}


{\it Dannie Heineman prize lecture, AAS Meeting, Washington D.C.,
January 2002}

\section{Introduction}

Observations of blue and uv surface brightnesses from young stars,
IR radiation from dust, and H$\alpha$ from HII regions tell us the
rates at which stars form in other galaxies (Kennicutt 1998a).
Observations of young stellar clusters reveal some of the
processes involved (Efremov 1995; Clarke, Bonnell, \& Hillenbrand
2000; Elmegreen et al. 2000), and observations of HI, CO, and
other gases, along with the associated dust structures, show how
these processes work (V\'azquez-Semadeni et al. 2000; Williams,
Blitz, \& McKee 2000).

These galactic-scale observations have led to two empirical laws for
star formation: a column density relation,
\begin{equation}
{\rm SFR/Area} \propto \Sigma^{1.4}
\end{equation}
(Schmidt 1959; Buat, Deharveng, \& Donas 1989; Kennicutt 1989,
1998b; Tenjes \& Haud 1991), and a column density threshold, for
both a fixed threshold,
\begin{equation}
\Sigma>\Sigma_{min} \sim 6 \; {\rm M}_\odot\;{\rm pc}^{-2}
\end{equation}
with gas column density $\Sigma$ (Gallagher \& Hunter 1984;
Skillman 1987; Guiderdoni 1987; Chiappini, Matteucci, \& Gratton et
al. 1997) and a variable threshold based on the Toomre (1964)
criterion:
\begin{equation}
\Sigma>\Sigma_{crit} \sim {{0.7c\kappa}\over{3.36 G}},
\end{equation}
corresponding to $Q\equiv c\kappa/\left(3.36 G\Sigma\right)<1.4$
for velocity dispersion $c$ and epicyclic frequency $\kappa$
(Quirk 1972; Zasov \& Simakov 1988; Kennicutt 1989).  Other
similar empirical laws have been suggested as well (e.g., Dopita
\& Ryder 1994; Prantzos \& Boissier 2000; Chiappini, Matteucci, \&
Romano 2001).

The column density relation holds for the main disks and inner
parts of galaxies as well as starburst regions in a continuous
power law (Kennicutt 1998b). It also works in the
Antennae galaxy for both general star formation and cluster
formation (Zhang, Fall, \& Whitmore 2001). The Toomre threshold
applies to normal spiral galaxies (Kennicutt 1989; Caldwell et al.
1992; Martin \& Kennicutt 2001), elliptical galaxies (Vader \&
Vigroux 1991), low surface brightness galaxies (van der Hulst et
al. 1993; Pickering et al. 1999), and starbursts (Shlosman \&
Begelman 1989; Elmegreen 1994a). The fixed column density
threshold is most evident in irregular galaxies 
(Taylor et al. 1994; Meurer et al.
1996; Hunter et al. 1998; van Zee et al. 1998; Hunter, Elmegreen,
\& van Woerden 2001).

There are a few interesting exceptions to these relations. O'Neil,
Bothun, \& Schombert (2000) and O'Neil, Verheijen, \& McGaugh
(2000) report low surface brightness galaxies with
$\Sigma>\Sigma_{crit}$ but not much star formation. Conversely,
the inner parts of M33 and NGC 2403 have $\Sigma<\Sigma_{crit}$
and normal star formation (Martin \& Kennicutt 2000), as does the
nuclear region of the S0/E7 galaxy NGC 4550 (Wiklind \& Henkel
2001).  Dwarf galaxies commonly form stars at column densities
that are a factor of 2 below what would be the threshold for
spiral galaxies (Hunter et al. 1998; van Zee et al. 1998). There
is also some concern that dynamical processes like spiral arm
generation should maintain $\Sigma\sim\Sigma_{crit}$ independent
of star formation (Fuchs \& von Linden 1998; Bertin \& Lodato
2001; Combes 2001).  Moreover, the threshold for instabilities
should not be $\Sigma_{crit}$ but $\sim0.4\Sigma_{crit}$ if the
combined stellar and gaseous fluids are considered (Orlova,
Korchagin, \& Theis 2002). Lower thresholds are also possible for
sub-populations of clouds with lower than average velocity
dispersions (Ortega, Volkov, \& Monte-Lima 2001), and for magnetic
disks with a Parker instability (Chou et al. 2000; Kim, Ryu, \&
Jones 2001; Franco et al. 2001). Thresholds for $\Sigma_{crit}$
based on the rate of shear rather than the epicyclic rate were
preferred by Pandey \& van de Bruck (1999).

Clearly the connection between star formation and disk stability
is more complex than originally envisioned in the standard
feedback scenario (Goldreich \& Lynden-Bell 1965).  Nevertheless,
there are several general implications of these empirical
relations that are not strongly dependent on the details of star
formation. The next section uses the column density thresholds to
show that gaseous self-gravity and a cool thermal state are
important precursors to star formation.  No particular mechanism
follows from these relations, however, because many cloud
formation processes have the same relations. Observations on
intermediate scales are discussed next. These data highlight the
importance of compressible turbulence as a scale-free cloud
formation process, but again fall short of pin-pointing the final
trigger for new stars.  The most telling observations come from
the morphologies of star-forming environments.  On small scales,
the common appearance of dense clusters inside comet-shaped clouds
adjacent to high-pressure HII regions or near the tips of isolated
filaments suggests that the final step in the star formation
process is a compression of pre-existing clouds by nearby young
stars and supernovae.

Considering this, the problem of star formation becomes one of
understanding how large and intermediate scale processes conspire
with local triggering events to give the empirical relations
discussed above. A possible solution is discussed in Sect.
\ref{sect:conspire} (see also Elmegreen 2002a).  
A related problem is whether such triggering
is important for all regions of star formation, including
starbursts, low surface-brightness disks, and the early Universe.
All of these
considerations should lead to a theory of star formation where the
rate can be predicted from first principles in all environments.
We are far from such a theory at the present time, but a few
suggestions for how it might go are made in Sect.
\ref{sect:theory}.

\section{Implications of the Empirical Laws for Star Formation on a Large Scale}
\subsection{The Dynamically-Based Surface Density Threshold}

The existence of a critical column density related to
$c\kappa/\left(3.36 G\right)$ implies that ambient self-gravity in
galaxy disks is important for star formation. This column density
threshold first arose in the context of spontaneous disk
instabilities, but it actually has a much broader applicability.

The instability model predicts that when $\Sigma
> c\kappa/3.36 G$ in a one-component stellar disk, spiral arms form easily (Toomre
1964, 1981; Goldreich \& Lynden-Bell 1965; Julian \& Toomre 1966;
Athanassoula 1984).  The gaseous parts of these arms collapse into
giant cloud complexes with a Jeans mass of
\begin{equation}
M_{Jeans}=\Sigma\left(\lambda/2\right)^2={{c^4}\over{
G^2\Sigma}}\sim10^6 - 10^7 \;{\rm M}_\odot \end{equation}
(Balbus 1988; Elmegreen 1994b; Kim \& Ostriker 2001) for wavelength
$\lambda=2c^2/\left(G\Sigma\right)$. Galaxies with numerous, short
and patchy arms show these collapse sites directly, one for each big
patch; they form randomly on the scale of the Jeans length ($\sim2$
kpc) and then get sheared into spiral arms at the same time as they make
stars. Galaxies with a few long and symmetric arms (Kuno et al. 1995) or
tidal arms (Rodrigues et al. 1999; Duc et al. 2000; Braine et al. 2001)
have comparable collapse sites, but they are located inside the arms,
strung out like beads with the same Jeans-length spacing. Models of
the first type are in Toomre \& Kalnajs (1991), Gerritsen \& Icke
(1997), Wada \& Norman (1999, 2001), Huber \& Pfenniger (2001) and
elsewhere. Models of the second type are in Elmegreen \& Thomasson (1993)
for spiral wave modes in normal galaxies, and Barnes \& Hernquist (1992)
and Elmegreen, Kaufman, \& Thomasson (1993) for tidal arms. Similar beads
of star formation line up around nuclear starburst rings (e.g., Sersic
\& Pastoriza 1965, 1967; Maoz et al. 1996; D.  Elmegreen et al. 1999;
Buta, Crocker \& Byrd 1999) and around the giant rings of collisional
ring galaxies (Bransford et al. 1998).  Presumably, these beads form by
the same types of gravitational instabilities.

In most nearby galaxies, the Jeans mass clouds at the top of this
collapse chain are seen as the largest coherent gas features
(Elmegreen \& Elmegreen 1983; Rand 1993; Kuno et al. 1995). They
are also present in the inner (Elmegreen \& Elmegreen 1987) and
outer (Grabelsky et al. 1987) Milky Way, enclosing hierarchical
clusters of molecular clouds and OB associations (Efremov 1995).

The extrapolation from this $\Sigma$-threshold behavior involving
gaseous self-gravity to the process of star formation in molecular
cloud cores is not obvious. The most straightforward scenario is
where gravitational instabilities sensitive to $\Sigma_{crit}$
make giant cloud complexes, as discussed above, and then these
complexes form molecules in their shielded cores, making giant
molecular clouds that dissipate turbulent energy, cool and
contract with increasing gravitational force until they form
stars. Such monotonic contraction probably happens some of the
time, but the complete picture seems to be more complex.

Part of the problem is that $\Sigma_{crit}$ also appears for other
processes of star formation, such as triggered collapse in
expanding shells (Elmegreen, Palou\v s, \& Ehlerov\'a 2002). When
$\Sigma<\Sigma_{crit}$, long-range triggered star formation
becomes inefficient as the expansion of shells is resisted by
Coriolis forces.  This makes it difficult for a shell to build up
enough material to become gravitationally unstable. The
probability that a shell triggers star formation, considering a
wide range of possible local conditions and rotation curves,
decreases from 0.6 to 0.1 as $\Sigma/\Sigma_{crit}$ decreases from
1.6 to 0.16. Thus $\Sigma>\Sigma_{crit}$ implies not only that
spiral arms can form and make clouds, but also that existing stars
can trigger other stars in shells with dimensions comparable to
the scale height.

$\Sigma_{crit}$ should have a similar role for star formation that
is triggered on large scales by turbulence compression.  This role
is most evident if we rewrite the Toomre $Q=c\kappa/\left(\pi
G\Sigma\right)$ parameter (for a gas $\pi$ replaces 3.36) in terms
of the ratio of the epicyclic radius at the local rms speed,
$R_{ep}=c/\kappa$, to the scale height of an isothermal layer,
$H=c^2/\left(\pi G\Sigma\right)$. Then $Q\equiv H/R_{ep}$. Large
$Q$ or small $\Sigma/\Sigma_{crit}$ correspond to turbulent eddies
that circle around in tight curls, narrower than a scale height,
because of the relatively large Coriolis force. Such eddies cannot
compress enough gas in their converging parts to make a Jeans
mass, considering that the ambient Jeans length is about equal to
the disk thickness.

The balance between self-gravitational and Coriolis forces
reflected in the ratio $\Sigma/\Sigma_{crit}$ affects any process
of {\it cloud} formation that involves moving and compressing a
gravitationally significant amount gas. Not every star formation
model should be sensitive to $\Sigma_{crit}$, however.  One where
random cloud collisions trigger stars should have no
$\Sigma_{crit}$ dependence. The fact that $\Sigma>\Sigma_{crit}$
in regions of star formation tells us only that disk gravity is
important for the first step in the star formation process, which
is cloud formation. These clouds may then form stars by a variety
of methods. Some will cool and condense by themselves, others will
collide, and still others will get compressed by nearby stars. To
investigate the mechanisms of star formation in more detail, we
have to look at individual regions. This is the topic of Sect.
\ref{sect:clusters}. Before this, the implications of the other
column density threshold, $\Sigma_{min}$, are considered.

\subsection{The Constant Surface Density Threshold}

The dynamically-based column density threshold, $\Sigma_{crit}$,
does not apply well to dwarf and irregular galaxies (Hunter \&
Plummer 1996; van Zee et al. 1997; Hunter, Elmegreen \& Baker
1998; Hunter, Elmegreen, \& van Woerden 2001). This may be because
there is no spiral-enhanced compression of gas in these galaxies,
and too little shear to pump turbulent energy into the gas after
spiral instabilities. There may be end-of-bar compressions in
dwarf galaxies (Roye \& Hunter 2000), but this would not involve
$\Sigma_{crit}$ either.  Thus several of the primary functions of
gaseous self-gravity in the star formation process for spiral
galaxies does not work well for dwarf galaxies. Nevertheless,
there is still some suggestion of an absolute minimum column
density for star formation in dwarfs, and this minimum seems to
apply to spirals as well.

The primary implication of a minimum column density for star
formation appears to be that the gas pressure must be high enough
to support a cool phase of HI clouds. The metallicity should also
be moderately high (Wolfire et al. 1995). The concept of multiple
interstellar thermal phases was introduced by Field (1965) and
Field, Goldsmith \& Habing (1969). Today, the two-phase model is
not considered to be the main driver of cloud formation because
most interstellar motions are supersonic and the thermal
instability always works at subsonic speeds, as measured in the
warm component. This means that gravitational instabilities,
expanding shells, and turbulence all make large clouds
faster than thermal instabilities (V\'azquez-Semadeni, Gazol, \&
Scalo 2000; Gazol et al. 2001). Nevertheless, the existence of a
cool phase in the atomic medium is a prerequisite for most star
formation, and the criterion for this phase is related to the
pressure, and therefore the gas column density, giving the
observed threshold effect.

The average pressure in an isolated galaxy disk is determined by
the weight of the gas layer in the gravitational potential of the
total disk mass that lies within the gas layer.  If the gas column
density is $\Sigma$, and the total column density inside the gas
layer including gas, stars, and dark matter is $\Sigma_{total}$,
then the midplane pressure is $P=(\pi/2)G \Sigma\Sigma_{total}$.
This relation comes from the following expressions: $P=\rho c^2$
for midplane gas density $\rho$ and velocity dispersion $c$,
$H=c^2/\left(\pi G\Sigma_{total}\right)$ for gas scale height $H$,
and $\Sigma=2\rho H$.  An approximation for $\Sigma_{total}$ in a
stars+gas disk is the expression
$\Sigma+\left(c_{gas}/c_{stars}\right)\Sigma_{stars}$, where the
last term accounts for the fraction of stars that are in the gas
layer (Elmegreen 1989a).   The gas pressure increases only
indirectly as the energy sources from supernova and stellar winds
increase: these sources affect $c$ most directly; $c$ affects $H$
through the vertical equilibrium, and $H$ affects the fraction of
the total stellar mass that is in the gas layer, which enters into
$P$.  The gas pressure can also increase in large regions if the
galaxy experiences a ram pressure from its flow through a hot
intergalactic gas.  Such an increase might have affected the
thermal balance at the leading edge of the LMC (Dickey et al.
1994).

The total pressure is important for the thermal state of the ISM.
When the pressure is high, a cool atomic phase of gas exists in
equilibrium where collisional cooling balances stellar heating
(Wolfire et al. 1995). If the pressure is very high, then only
this cool atomic phase exists, in addition to cold and possibly
warm molecular phases and a warm or hot ionized phase. The warm
diffuse molecular phase is not important for CO in the Solar
neighborhood but it is moderately important in the inner Milky Way
and very important in starbursts and other active regions with
high pressures (Aalto et al. 1995; Wilson, Howe, \& Balogh 1999;
Smith, et al. 2000; H\"uttemeister et al. 2000; Mao et al. 2000;
Curran et al. 2001; Israel \& Baas 2001; Rodr\'iguez-Fern\'andez
et al. 2001). For intermediate pressures, as in the Solar
neighborhood, both the warm and cool phases of atomic gas co-exist
and there is warm $H_2$ gas in diffuse clouds without much CO
emission. Where the pressure is low (and usually the metallicity
is low at the same place), only the warm phase can exist for
atomic gas, and then there are few cool diffuse atomic clouds and
few molecular clouds. The molecules also tend to be confined to
the far inner regions of self-gravitating clouds when the ambient
pressure and metallicity are low (Lequeux et al. 1994), and there
is a greater abundance of very cold ($T\sim10-40$ K) atomic clouds
in the place of normal molecular clouds (Dickey et al. 2000).  The
minimum pressure for the existence of cool diffuse atomic clouds
corresponds to a minimum gas column density threshold of about
$\Sigma> 6$ M$_\odot$ pc$^{-2}$, depending slightly on the
metallicity and radiation field (Elmegreen \& Parravano 1994).
This presumably gives the observed $\Sigma_{min}$ threshold for
star formation.

Nearby spiral galaxies clearly reveal this column density
threshold.  Braun (1997) noted how the outer parts of all the
spiral galaxies he considered make a transition to a smooth warm
phase just beyond the optical radius, $R_{25}$.  For NGC 2403, the
fractional mass in the form of cool HI drops from 90\% to 20\% as
the radius increases from $\sim0.8$R$_{25}$ to $\sim1.6$ R$_{25}$;
beyond $\sim2$R$_{25}$, this fraction is less than 10\%.
Similarly, the Sagittarius dwarf irregular galaxy has two
components of HI that may be separated by their linewidths;  star
formation occurs only near the cool component (Young \& Lo 1997b).
The same is true in other dwarf galaxies (Young \& Lo 1996, 1997a) and in
NGC 2366 (Hunter, Elmegreen, \& van Woerden 2001).  In the main
disk of the LMC, the proportion of HI in cool diffuse clouds is
the same as it is in the Solar neighborhood, but there is an
increase in this proportion, and a possible decrease in cloud
temperature, toward the 30 Dor region (Dickey et al. 1994; Mebold
et al. 1997) and LMC4 (Marx-Zimmer et al. 2000), where the
pressures are high. There is also cool HI in the LMC tidal bridge,
suggesting moderately high pressures there too (Kobulnicky, \&
Dickey 1999). The SMC differs though, having a lower fraction of
HI in a cool diffuse form, and lower temperatures in that form,
presumably because of its lower pressure and metallicity (Dickey
et al. 2000). The cool HI fractions in the main disks of M31 and
M33 are similar to that in the Milky Way (Dickey \& Brinks 1993).

The outer parts of both spiral and dwarf irregular galaxies are so
warm that their total HI linewidths are comparable to the thermal
speed.  No source of turbulence is needed for the outer parts of
galaxies in this case. There is usually enough stellar light from
the inner disk to keep the atomic gas warm (Elmegreen \& Parravano
1994), so the velocity dispersion stays moderately high even as
the column density drops. This means that any tendency for stellar
energy feedback and gravitational instabilities to regulate
$\Sigma\sim\Sigma_{crit}$ through star formation and spiral arm
formation should turn off when $\Sigma<\Sigma_{min}$.

A cool phase of atomic gas is necessary for star formation because
cool diffuse clouds are the first step in the transition from the
average ISM to the dense molecular clouds where stars form. The
average ISM usually has a gas density comparable to the critical
tidal density, which is $=-3\Omega R/\left(2\pi G\right)
d\Omega/dR\sim 1 $ cm$^{-3}$ locally, for galaxy angular rotation
rate $\Omega$ and galactocentric radius $R$. This density is
always too low to form stars directly. The density has to be high
enough that the dust column out to the nearest bright star is
opaque to the star's uv light.  Then the gas can cool and build up
molecules that lower the temperature even more with their
low-excitation rotational transitions. Eventually the temperature
gets so low ($\sim10$ K), and the density so high in a
near-pressure equilibrium, that the self-gravitational energy
density in turbulence-compressed regions overcomes the thermal
energy density. Collapse to stars follows if the local shear and
turbulent energy densities are also low.  Note that it is not
enough for gravity to dominate thermal pressure ($M>M_{Jeans}$) in
a turbulent region: star formation also requires that the clump
last for a collapse time before the external turbulent flows
destroy it.  Excessive turbulence could conceivable prevent star
formation if the motions continuously force the gas to break up
into pieces that are smaller than a thermal Jeans mass (Padoan
1995).

Without the first step of diffuse cloud formation, ambient
self-gravity, spiral wave shocks, supernovae, turbulence, and
other large-scale disturbances cannot make the gas dense enough to
become opaque and form cold molecular clouds. Only if the ambient
ISM already has diffuse clouds, as in the main disks of spiral
galaxies, or if a pressure disturbance in a marginally warm ISM is
strong enough to induce the transition to a cool diffuse state,
can star formation proceed.  Dense clouds also form their
molecules faster if the diffuse clouds that make them are mostly
molecular $H_2$ too (Pringle, Allen \& Lubow 2001).  The outer
disks of galaxies are generally warm and without much star
formation, but sometimes a spiral arm can act as a pressure source
and make enough cool gas to start the process off (e.g., see
Ferguson et al. 1998; LeLi\`evre \& Roy 2000). This illustrates
the old model by Shu et al. (1972), generalized recently by Koyama
\& Inutsuka (2000). Better observations of the most distant
star-forming regions in our Galaxy may elucidate the possible
pressure sources (Kobayashi \& Tokunaga 2000).

\subsection{The Schmidt Law}
\label{sect:schmidt}

The other empirical law of star formation, the power-law
dependence of the star formation rate on the total column density,
indicates that cloud and star formation usually occur at the local
dynamical rate averaged over a large area.  If we write this law
as
\begin{equation}
{\rm SFR/Area} \sim
\epsilon\Sigma\omega\label{eq:schmidt}\end{equation} for
efficiency $\epsilon\sim$few percent and rate
$\omega\sim\left(G\rho\right)^{1/2}\propto\Sigma^{1/2}$ for
constant $H$, then the $\Sigma^{1.4}$ law follows approximately
(Madore 1978). Most of the exponent in this relation is from the
available gas, which contributes $\epsilon\Sigma$ to the star
formation rate; $\omega$ is the conversion rate of this gas into a
dense form. The use of a dynamical rate for $\omega$ does not
imply that gravitational forces are directly involved. The
dynamical rate is also about equal to the turbulence crossing rate
over a scale height ($c/H=\left[2\pi G\rho\right]^{1/2}$) and it
is the inverse of the collapse time for large expanding shells
with modest overpressures (i.e., for low Mach numbers -- see
Elmegreen, Palou\v s, \& Ehlerov\'a 2002).

Models of galactic evolution using an equation like equation
\ref{eq:schmidt} give a reasonable agreement with the observed
radial dependence of star formation (Wang \& Silk 1994). The
expression is sometimes multiplied by another term proportional to
the galactic orbital rate, as if a spiral shock were involved too
(Wyse \& Silk 1989; Prantzos \& Boissier 2000).  This is done to
better reproduce the metallicity gradient, which is not steep
enough from equation \ref{eq:schmidt} alone. However the
additional factor is not observed directly in star formation
studies, and the steep gradients could result from other things,
such as global gas accretion (e.g., Ferguson \& Clarke 2001 and
references therein).

The Schmidt law is not very model-dependent because the detailed
physics of the star formation process is confined to a relatively
weak dependence on the density or column density, i.e., in the
$\omega$ term.  This means that nearly any model with a dynamical
time scale ($\propto \left[G\rho\right]^{-1/2}$) can give it, as
can other models, such as propagating star formation (Sleath \&
Alexander 1995) and cloud collisions (Tan 2000).

The Schmidt law is inconsistent with the 
recent suggestion that the efficiency of
star formation,
as given by the rate per unit gas mass
and equal to the inverse of the gas consumption time,
is about constant in
a variety of environments, including normal
galaxies (Rownd \& Young 1999; Boselli, Lequeux, \& Gavazzi 2002),
the central regions of early-type galaxies (Inoue, Hirashita, \&
Kamaya 2000), and the starbursting antennae galaxy (Gao et al.
2001).  This would imply that SFR/Area $\propto\Sigma$ only, without
the $\omega$ factor. 
The discrepancy between these two observations has not been explained. 
We return to the
Schmidt law in Section \ref{sect:theory}.

\section{Star Formation on Intermediate Scales}

Ambient interstellar gravity and a cool phase of atomic gas are
important for star formation, but they do not work alone. Clouds
formed by ambient gravitational instabilities should have a
characteristic size and mass at first, and possibly a regular
structure too.  There are two characteristic sizes, the Jeans
length, $c^2/\left(\pi G\Sigma\right)$, and the Toomre (1964)
length, $2\pi G\Sigma/\kappa^2$. The first arises from the balance
between pressure and self-gravity and appears in the separation
between clouds along spiral arms (Kuno et al. 1995).  The second
results from the balance between Coriolis forces and self-gravity
and appears in the separation between stellar arms.  This
regularity differs from the morphology of clouds and star
formation on intermediate and small scales, where scale-free and
fractal structures are seen.

The earliest indication that the ISM is scale free came from the
mass spectra of clouds and clusters (e.g. Field \& Saslaw 1965).
The cloud mass spectrum is a power law below the ambient Jeans
mass (e.g., Heyer, Carpenter, \& Snell 2001), and the cluster mass
spectrum is a power law for the same physical scales (Battinelli
et al. 1994; Comeron \& Torra 1996; Elmegreen \& Efremov 1997;
Feinstein 1997; McKee \& Williams 1997; Oey \& Clarke 1998; Zhang
\& Fall 1999).

These two power laws are usually seen only in a piecewise fashion
for the gas, covering 2 orders of magnitude in cloud mass at most.
When large and small pieces are fit together in any one region,
using different angular resolutions, the mass range can be
extended (Heithausen et al. 1998). Single maps show only a factor
of $\sim100$ in mass because the mass scales as the square of the
size (Larson 1981), and cloud-finding routines recognize only
factor of $\sim10$ in size as a result of selection effects
(Elmegreen \& Falgarone 1996). The minimum cloud size is always
several times the telescope beam width (Verschuur 1993), and the
maximum cloud size is just big enough to contain recognizable
substructure, at which point a decision is usually made to divide
the bigger cloud into its parts and call them separate clouds. The
largest gas concentrations in big maps are usually not included in
the derived mass spectra as single clouds but only as collections
of smaller clouds.

A better way to see a wide range of scale-free structure is with
Fourier transform power spectra, which, for wide fields in the
Milky Way (Crovisier \& Dickey 1983; Green 1993; St\"utzki et al.
1998; Dickey et al. 2001), whole galaxies (Stanimirovic, et al.
1999; Elmegreen, Kim, \& Staveley-Smith 2001), and galactic nuclei
(Elmegreen, Elmegreen, \& Eberwein 2002) show no characteristic scale between
the size of the beam and the size of the map, except possibly for
the line-of-sight galaxy thickness if that is resolved (Elmegreen
et al. 2001; Padoan, et al. 2001a).  The Delta variance technique
(St\"utzki et al. 1998; Zielinsky \& St\"utzki 1999; Bensch,
St\"utzke, \& Ossenkopf 2001) and spectral correlation function
(Rosolowsky et al.  1999) are other ways to see scale-free
structure that have some advantages over Fourier transform power
spectra if there are sharp map boundaries and spectral line
information, respectively.  Scale-free variations for smooth
intensity distributions are called multifractal because the
fractal dimension varies between the peaks and the valleys (e.g.,
Vavrek 2001). IRAS intensity distributions of the local dust
emission were shown to be multifractal by Chappell \& Scalo
(2001a).  The edges of clouds (Dickman, Horvath, \& Margulis 1990;
Falgarone, Phillips, \& Walker 1991) and galaxies (Westpfahl et al
1999) are fractal, which means their irregularities are scale-free
(Mandelbrot 1983).

Stars form in this scale-free gas by making scale-free clusters
and aggregates.  This means that star fields are hierarchical if
their ages are less than a crossing time (Feitzinger \& Galinski
1987; Gomez et al. 1993; Efremov 1995; Elmegreen \& Efremov 1996;
Battinelli, Efremov \& Magnier 1996; Harris \& Zaritsky 1999;
Testi et al. 2000; Elmegreen 2000; Heydari-Malayeri et al. 2001;
Pietrzynski et al. 2001; Elmegreen \& Elmegreen 2001; Zhang,
Fall, \& Whitmore 2001). Older regions may have power law
structures too, but for different reasons (Larson 1995; Simon
1997; Bate, Clarke \& McCaughrean 1998; Nakajima et al. 1998;
Gladwin et al. 1999).

There is no characteristic length, like an ISM Jeans length, in
the distribution of young stars. Even when the maximum size of a
star complex looks like a Jeans length (Elmegreen et al. 1996), it
may really be the maximum likely length from sampling statistics
(Selman \& Melnick 2000).  This is because gas concentrations that
begin at the Jeans length cascade both downward and upward in
scale from the combination of turbulence, self-gravity, and shear,
leaving little signature of their initial structure.  An exception
occurs where spiral instabilities do not operate, as in the
centers of density-wave spiral arms; there the shear is low or
negative and the Jeans mass condensations live only a short time
before they flow into the interarm region.

The scale-free nature of interstellar gas is the imprint of
turbulence (e.g., Falgarone \& Phillips 1990; Scalo 1990; Lazarian
1995; Goldman 2000), or turbulence combined with non-linear
self-gravitational instabilities (Fuchs \& von Linden 1998;
Semelin \& Combes 2000; Crosthwaite, Turner, \& Ho 2000; Bertin \&
Lodato 2001; Wada \& Norman 2001; Huber \& Pfenniger 2001; Vollmer
\& Beckert 2002; Chavanis 2002). Chaotic structures can also come
from the thermal instability (Elphick, Regev, \& Spiegel 1991).

An important point about turbulence is that smaller scales have
smaller internal velocity dispersions. This means that the Jeans
length decreases along with the physical scale, making the ratio
of gravitational to turbulent energy densities somewhat constant
(Larson 1981). For clouds that are defined out to a fixed opacity
threshold, as in CO surveys, the pressure boundary condition
breaks this constancy and makes the ratio of cloud mass to Jeans
mass systematically decrease with scale.

The luminosity-based CO mass is $M_{CO}\propto TcR^2$ for velocity
dispersion $c$, radius $R$, and excitation temperature, T, and the
Jeans mass is $M_{J}\propto c^2R$, so $M_{CO}/M_J\propto TR/c$,
which is the crossing time for constant $T$.  For gravitating
clouds, the column density scales with external pressure,
$P\sim0.1GM^2/R^4$, giving $c\propto R^{1/2}$ with the virial
theorem. For non-self-gravitating clouds, $c$ and $R$ have the
same relation from turbulence.  As a result, $M_{CO}/M_J\propto
TR^{1/2}$, so small CO clouds are systematically less
self-gravitating than large CO clouds at the same $T$. The
threshold is at about $10^4$ M$_\odot$ for the FCRAO outer galaxy
CO survey (Heyer, Carpenter \& Snell 2001). It is larger
($\sim10^6$ M$_\odot$) for collections of CO clouds inside giant
molecular associations or HI superclouds (Inoue \& Kamaya 2000)
and smaller ($\sim10^3$ M$_\odot$) for the cores of CO clouds
(Bertoldi \& McKee 1992; Falgarone, Puget, \& P\'erault 1992) and
outer galaxy CO clouds (Brand 2001). These differences in mass at
the virial threshold $M_{CO}=M_J$ imply that the coefficient
$\alpha$ in the size-linewidth relation, $c=\alpha R^{1/2}$,
varies among the different surveys, not as a function of scale but
as a function of molecule type, pressure, or perhaps survey
sensitivity.  Numerical simulations of turbulence that consider
these relations in more detail are in Vazquez-Semadeni,
Ballesteros-Paredes \& Rodriguez (1997).

This is only one example of the many selection effects that can
result from cloudy-model interpretations of turbulent structures
and motions.  Scalo (1990) recognized some of these issues at an
early stage.  Other problems for the gas are the appearance of
false clouds from velocity crowding on the line of sight
(Ballesteros-Paredes, V\'azquez-Semadeni, \& Scalo 1999; Lazarian
\& Pogosyan 2000; Pichardo et al. 2000; Ostriker, Stone, \& Gammie
2001; Lazarian et al. 2001), and the false separation of ``cloud''
types (molecular, atomic, self-gravitating, diffuse) when the
physical distribution of the gas is more of a continuum.

Analogous selection effects result for stellar distributions after
conceptually forcing a discrete ``cluster'' model onto what is
really a more complicated pattern. Terms like clusters, OB
associations, stellar aggregates, and star complexes are different
scales inside a hierarchy of self-similar structures.  Dense
clusters probably form by processes that are very similar to those
involved with giant star complexes, even though dense clusters
look very different in their relaxed state. These differences in
morphology as a function of scale result from three effects that
are unrelated to the star formation process itself: (1) Massive
regions sample further into the high mass tail of the initial
stellar mass function, forming O-type stars readily. (2) Large
regions form stars long enough to make supernovae within a
dynamical time scale, causing severe cloud dispersal and an
inability to form stars with a high efficiency; the result is an
unbound collection of stars after the gas leaves. (3) Galactic
tidal forces shear away the remains of large-scale star formation
because the average density is low.

The turbulence scaling laws cause these three effects. As the size
of a region increases, the time scale for star formation increases
approximately as the square root of size and the average densities
of the gas and clusters decrease approximately inversely with
size. Most of the morphological differences between bound open
clusters and loose star complexes seem to result from these
initial scale-dependent differences in the gas. Power-law mass
functions for galactic clusters, which sample only the densest
cores of the multifractal gas (Elmegreen 2002b) and for star
complexes, which sample the largest scales (D. Elmegreen \& Salzer
1999) follow from the fractal distribution of the gas.

The apparent similarity in star formation processes over a wide
range of scales becomes more obvious at very high pressures, where
the giant unbound star complexes that form in normal galaxies
change to become morphologically indistinct from bound galactic
clusters, i.e., they form super star clusters or young globular
clusters with the mass of a complex but the density of a galactic
cluster. This change occurs because ambient pressures exceeding
$\sim10^7$ k$_{\rm B}$ lead to turbulent crossing times that are
less than the supernova time of an O-type star in a region
containing $10^6$ M$_\odot$ or more (Elmegreen \& Efremov 1997;
Ashman \& Zepf 2001).  Another change may occur if the cluster
environment is so dense that coalescence and protostar
interactions affect the IMF (Bonnell et al. 2001).

Self-similarity in star formation is also apparent when viewing
extremely young embedded clusters or clusters of pre-stellar objects.
There the source distribution is still hierarchical, not relaxed or
isothermal like the old notion of a cluster (Motte, Andre, \& Neri 1998;
Testi et al. 2000). The relaxation occurs after individual stars form
and begin to interact with gas and other stars.

Hierarchical structure in young clusters implies that star
formation is faster than an orbit time; otherwise the stars would
mix and scatter (Ballesteros-Paredes, Hartmann \&
V\'azquez-Semadeni 1999; Elmegreen 2000; Carpenter 2000; Yamaguchi
et al. 2001a). What defines a cluster before one dynamical time is
the high density of cores, protostars, and stars that form in high
density gas. Viewed from a perspective that includes a wide range
of scales, this clustering property is not special or limited to
small sizes. In an embedded cluster, the mean separation between
stars is proportional to the inverse cube root of the average
density, by mass conservation. There is usually no evidence for a
characteristic length from discrete physical processes until the
binary star separation is reached (Larson 1995). When the gas is
mostly used up the cluster begins to disperse (Kroupa, Aarseth \&
Hurley 2001). Perhaps 90\% of all stars younger than a few tenths
of a million years occur in dense embedded clusters (Carpenter
2000), but only $\sim50$\% of them remain after $\sim10$ My
because of this systematic dispersal (Battinelli \&
Capuzzo-Dolcetta 1991; Yamaguchi et al. 2001a).

Hierarchical structure breaks down if there are regular filaments
or sheets in the gas that are controlled by external forces. Such
structures condense into regularly-spaced globules as a result of
gravitational instabilities (Miyama, Narita, \& Hayashi 1987;
Bastien et al. 1991). Regular structure is evident in the Orion
core, for example (Vannier et al. 2001). The power law slope of
autocorrelated structures also changes on small scales when
self-gravity becomes strong because then the dense cores that form
are not transient turbulent structures (Ossenkopf, Klessen, \&
Heitsch 2001).

The correlation between size and crossing time that is well-known
for turbulent gas is visible also in the stars as a correlation
between the size of a region and the duration of the star
formation event there. This stellar correlation ranges between
$\sim1$ pc and $\sim1000$ pc, with larger regions taking longer to
evolve but each region forming stars completely in several local
dynamical time scales (Elmegreen \& Efremov 1996; Efremov \&
Elmegreen 1998; Battinelli \& Efremov 1999; Elmegreen 2000;
Heydari-Malayeri et al. 2001). This means that OB associations and
their GMCs appear to have a characteristic size and mass, but only
because the selection of a region by the presence of O-type stars
defines the age and therefore the scale. Selection by the presence
of Cepheid variables and red supergiants defines a much larger
characteristic scale because the relevant timescale is longer
(Efremov 1995).

Selection effects also bias our inference of a maximum scale for
star formation.  There is a maximum scale for OB associations in
the Milky Way at which there is a sudden drop in the luminosity
function (McKee \& Williams 1997). There is no physical limit to
star formation at this scale, however; the clustering of young
stars continues up to star complexes and flocculent spiral arms
(Elmegreen \& Efremov 1996). These larger regions form stars for
longer times and therefore contain several separate OB
associations as subregions (e.g., Comeron 2001). They would not be
called single OB associations. Thus the count of OB associations
in a whole galaxy, like the count of clouds in the previous
discussion, has a size limitation above which the physical
structures that are present tend to be resolved out, subdivided,
and then ignored as distinct entities. Autocorrelation studies for
young star fields find no characteristic feature on the scale of
an OB association, only a systematic weakening of the correlated
power up to the point where a statistically small number of
distinct units remains (D. Elmegreen \& Salzer 1999; Selman \&
Melnick 2000; Zhang, Fall, \& Whitmore 2001; Elmegreen \&
Elmegreen 2001).

In summary, the morphology of star formation on intermediate
scales points to the strong influence of turbulence in compressing
the gas and defining its dynamical time. These intermediate scales
range from the Jeans and Toomre lengths (as well as the disk
thickness) at the upper end, down to the scale of significant
cloud erosion and distortion inside HII regions.  This morphology
suggests that in spite of the empirical correlations mentioned in
Section I, which point most directly to gravitational
instabilities as the cause of star formation, the actual dynamics
involved is more closely related to transient compression in a
turbulent fluid (Elmegreen 1993; Myers \& Lazarian 1998; Klessen,
Heitsch, \&  Mac Low 2000; Williams \& Myers 2000; Heisch, Mac
Low, \& Klessen 2001; Klessen 2001; Padoan et al. 2001b).  The
processes that make the 
empirical laws serve mostly to regulate this turbulence.
Self-gravity is necessary to pump turbulent energy into the gas on
large scales through spiral instabilities, and it is also
necessary to make self-gravitating clouds in the compressed
regions that result from other processes. This link between
turbulence and self-gravity on the large scale, defined by the
condition $\Sigma\sim\Sigma_{crit}$, also ensures a basic equality
between gravitational and turbulent energy density on small
scales, where dense cores eventually form clusters. That is, the
compressed regions in a turbulent fluid become self-gravitating
and last for a collapse time if the larger scale gas around
them is also self-gravitating (Elmegreen 1993). The
turbulence-gravity link also shows up in the conversion rate from
ambient gas to star-forming gas, which occurs on both the
dynamical time and the turbulence crossing time ($\omega$ in
equation \ref{eq:schmidt}) when $\Sigma\sim\Sigma_{crit}$.

\section{Star Formation in Dense Clusters}
\label{sect:clusters}

The previous sections cited observations on large and intermediate
scales which suggest that self-gravity in the interstellar gas
forms $M_{Jeans}=10^7$ M$_\odot$ cloud complexes and drives
turbulence in and around these complexes as a result of
swing-amplified spiral instabilities if there are no stellar arms,
and beading instabilities inside stellar arms if there is a global
density wave. The observations also suggest that this turbulence
expands to larger scales by the swing amplifier and cascades to
smaller scales by non-linear hydrodynamics.  Pressure fluctuations
from supernovae and other stellar processes add to this turbulence
on intermediate to small scales. As a result, the gas is
compressed into a fractal network with marginally opaque clouds if
there is a cool atomic or molecular phase available. The structure
has a wide range of scales, and star formation follows inside the
cold molecular cores of the most strongly self-gravitating parts.

Understanding the last part of this scenario, how star formation
actually begins, requires observations of young clusters and not
just interstellar gas structures.  The biggest clue is that most
embedded clusters in the solar neighborhood are adjacent to HII
regions excited by slightly older clusters. This is true for Orion
(Lada et al. 1991; Dutrey et al. 1991; Reipurth, Rodriguez, \&
Chini 1999; Coppin et al. 2000), the Rosette nebula (Phelps \&
Lada 1997), Perseus OB2 (Sancisi et al. 1974; Sargent 1979),
Ophiuchus (de Geus 1992), Sco-Cen (Preibisch \& Zinnecker 1999),
the Trifid Nebula (Lefloch \& Cernicharo 2000), W3/4/5 (Thronson,
Lada, Hewagama 1985; Kerton \& Martin 2000), M17 (Thronson \& Lada
1983) and dozens of other regions, as reviewed in Elmegreen (1998)
and Elmegreen et al. (2000). The whole Perseus arm from
$l=106^\circ$ to $140^\circ$ has most of its IRAS sources in
clusters that are at the tips of cometary clouds and partial
shells around HII regions (Carpenter, Heyer \& Snell 2000;
Bachiller, Fuente, \& Kumar 2002).

In all of these cases, the embedded clusters are young enough to
have been triggered by the pressures of the adjacent HII regions.
They are not just highly obscured parts of the same clusters that
excite the HII regions.  For this reason, most clusters (and
therefore most young stars) look triggered in pre-existing clouds
by the sudden application of a high external pressure. Yamaguchi
et al. (1999) estimate that several tens of percent of all star
formation in the inner Galaxy is triggered by adjacent HII
regions.  The same morphology is found in other galaxies,
including the LMC where giant shells (Goudis \& Meaburn 1978; Kim,
et al. 1999; Yamaguchi et al. 2001b), the 30 Dor region (Walborn
et al. 1999) and other regions (Heydari-Malayeri et al. 2001) have
high-pressure triggering. Yamaguchi et al (2001c) estimate again
that a substantial fraction of star formation in the LMC is
triggered in this way. Galaxies that are much further away do not
yet have high resolution infrared observations to show embedded
young clusters, but still the juxtaposition of high pressure and
star formation is evident in the form of giant star-forming shells
(e.g., Brinks, Braun, \& Unger 1990; Puche et al. 1992; Stewart \&
Walter 2000; Stewart et al. 2000). Evidence for triggering in a
dwarf starburst galaxy was given by MacKenty et al. (2000).

Smaller regions of star formation, such as Taurus and Serpens, do
not sample the IMF far enough into the high mass tail to make the
massive stars that excite HII regions. As a result, star formation
looks spontaneous there, although the clouds themselves could have
formed in large-scale turbulent flows (Ballesteros-Paredes,
Hartmann \& V\'azquez-Semadeni 1999). However, the time scale for
energy dissipation and spontaneous star formation in these regions
is longer than the time between stray supernovae, so the final
compression into dense cores could have been triggered after all.
The Taurus clouds, for example, have a wind-swept appearance (Fig.
2 in Hartmann, Ballesteros-Paredes \& Bergin 2001), suggesting
that explosions or high speed flows from the east compressed and
triggered the main star-forming core. Core D in TMC1 (Hirahara et
al. 1992), with its peculiar molecular abundances suggesting
extreme youth (Hartquist, Williams, \& Viti 2001), is in a ridge
near the head of this wind-swept structure, as if it were just
compressed. Similarly, IC 5146 is a long filamentary cloud with a
molecular core and young cluster at the tip (Lada, Alves, \& Lada
1999). Filaments like this have a high cross section for stray
supernovae and could be triggered easily at one end or the other
without much residual evidence.

Sequential triggering is pervasive for three reasons: the
hierarchical structure of the gas implies that most clusters have
neighboring clouds to compress; the inefficiency of star formation
ensures there is always residual gas that can be triggered
further; and the spontaneous processes inside a non-star-forming
cloud usually take longer than the pressure fluctuations outside
the cloud. This latter inequality, mentioned also in the previous
paragraph, warrants more discussion.

The internal dynamical processes in a turbulent medium are usually
quicker than the external dynamical processes because the
turbulent crossing time decreases with scale. However, turbulence
dissipates significantly on only one crossing time (Stone,
Ostriker, \& Gammie 1998; MacLow et al. 1998), so the ISM must be
stirred more frequently than the crossing rate if it is to remain
turbulent. If turbulent energy input is pervasive, as in the
reaction-advection model by Chappell \& Scalo (2001b), then for
nearly any scale in the midst of these stirring motions, the
pressure fluctuations coming from outside a region should dominate
the spontaneous pressure fluctuations happening inside the region.
On scales that are much smaller or larger than the typical scales
at which turbulent energy is deposited, spontaneous evolution
should dominate triggered evolution as dense self-gravitating
regions contract toward denser states.  Inside a cluster-forming
core, for example, which is somewhat shielded from common ISM
disturbances, the self-gravitating clumps and protostellar
condensations should be able to evolve independently toward stars.
Also on galactic scales that are too large for common stellar
energy sources to perturb in a coherent fashion, gaseous
structures like spiral arms grow spontaneously from
swing-amplified noise. But over the range of scales between these
limits, including the scales for resolved atomic and molecular
clouds in galactic surveys, the environment outside a cloud
boundary is hostile and often more influential than the
environment inside it before star formation begins. For this
reason, most diffuse and molecular clouds should be triggered into
forming their first generation of stars by random supernova and
other specific events. Subsequent generations in the same cloud
should be triggered sequentially because the pressure from the
first generation will most likely dominate the environmental
pressure later.

This balance between internal and external triggers could shift in
young starburst regions, which are usually places where the
epicyclic frequency and average dynamical rate are very high. If
the dynamical time is much less than the lifetime of an O-type
star, which is 3 My, requiring ambient ISM densities exceeding
$\sim100$ cm$^{-3}$, then spontaneous star formation can proceed
for quite a while (in a relative sense) before supernova begin.
There should be proportionally less triggering because of this.
Also in these regions, high pressures give self-gravitating clouds
high velocity dispersions.  When this dispersion exceeds $\sim10$
km s$^{-1}$, HII regions are born in near-pressure equilibrium so
they do not expand, compress gas, and trigger more star formation.

Similarly, if the star formation rate is extremely low, as in the
outer parts of galaxies or during quiescent periods in irregular
galaxies, then supernova and other stellar pressures can be rare
and pre-existing clouds will not be triggered. Slow spontaneous
instabilities then dominate the star formation process, forming
giant cloud complexes with dense cool cores if the internal
pressure is high enough.

\section{A Star-Formation Conspiracy}
\label{sect:conspire}

The similarities between the empirical laws of galactic star
formation and the threshold and dynamical properties of
gravitational instabilities in galaxy disks do not uniquely
identify these instabilities as the cause of star formation.
Direct compression of pre-existing clouds seems to cause most of
the local star formation, while the time and spatial correlations
for both gas and young stars suggests that turbulence is involved
too. How do these three unique processes work together in such a
seamless fashion?

The primary role of the dynamical threshold for column density,
$\Sigma_{crit}$, is to determine where Jeans-mass clouds form as
self-gravity overcomes Coriolis and pressure forces. The localized
collapses appear as small spiral arms in the gas and stars, or as
beads of clouds in the large stellar arms. The instabilities
themselves do not necessarily lead to star formation, although
they can, but the motions they induce drive turbulence that
compresses the gas further. This secondary compression covers a
wide range of scales below the Jeans length and stretches to even
greater lengths because of shear. The cascade downward quickly
produces very small structure in the cool component of the ISM.
The minimum column density threshold ensures that there is such a
cool phase available.  All of this happens continuously in the
main disks of galaxies. The turbulent cascade is relatively fast
compared to other processes so we rarely see a perfectly smooth
ISM on the scale of the Jeans length.  In the outer parts of
galaxies, or wherever the column density is less than the minimum
for a cool phase, the ISM can be smooth.

Gas motions and cloud formation that are driven on large and
intermediate scales by ISM gravity and turbulence could initiate
star formation if given the chance, but in the main disks of
galaxies there is usually so much activity that external pressure
fluctuations tend to initiate star formation first.  This is true
whether the cloud was made by random turbulent motions in the next
larger level of the hierarchy or by pressure-driven accumulations
around an existing star formation site.  A cloud is also more
likely to be externally triggered than to form stars on its own
whether it lies in an OB association and has a short internal
dynamical time (because of the high environmental pressure), or
lies in the intercloud medium and has a long dynamical time scale.

The implication of these considerations is that a fractal ISM, set
up by one sequence of processes, is continuously rattled and
pounded by pressure fluctuations from another sequence of
processes.  That is, star formation is often triggered by stellar
pressures in clouds that turbulence, self-gravity, and other
stellar pressures make. This interpretation explains the empirical
laws mentioned in the introduction, it explains the scale-free
structure of gas and young star fields, the origin of most
turbulence and its connection with gravity, and the common
appearance of young clusters in comet-shaped clouds and at the
edges of high pressure HII regions.  It also explains why the star
formation rate in dwarf irregular galaxies, which provide a good
test for local processes because they do not have spiral waves,
scales better with the blue surface brightness than with either
the gas column density or the threshold $\Sigma_{crit}$ (Hunter,
Elmegreen, \& Baker 1998; Brosch, Heller, \& Almoznino 1998). The
blue surface brightness traces the existing stars, and the
pressures from these stars initiate star formation in the
turbulent clouds that happen to be nearby.

The size-duration correlation in young star fields remains to be
explained. This correlation resembles the size-crossing time
correlation for turbulence, making it look like turbulent motions
hold clouds in place externally while they form stars internally
on the dissipation time, which is the internal turbulent crossing
time.  Such a process may actually apply to individual stars in
the interiors of clouds (e.g. Williams \& Myers 2000), but the
morphology of such compression contrasts with the cometary and
shell-like morphology of star-forming regions on larger scales,
which are often adjacent to HII regions or ambient pressure
excursions as discussed above.

There is another aspect of triggered star formation that seems to
have some bearing on the size-duration correlation.  The time
scale for triggering by high pressure events is usually comparable
to the dynamical time scale in the external, low-pressure medium.
For example, the collapse time in compressed layers and shells,
while depending only on the internal layer density (Vishniac 1983;
McCray \& Kafatos 1987), turns out in practice to depend mostly on
the external (pre-shock) density at the relatively late time when
embedded clusters form.

This external density dependence has been demonstrated in several
ways. First, gravitational instabilities on the time scale of the
internal density are relatively fast and they accompany the
kinematic instabilities that result from shock curvature
(Doroshkevich 1980; Welter \& Schmid-Burgk 1981; Vishniac 1983,
1994; Nishi 1992; Yoshida \& Habe 1992; Kimura \& Tosa 1991, 1993;
Lubow \& Pringle 1993; Strickland \& Blondin 1995; Garcia-Segura
\& Franco 1996). These kinematic instabilities drive strong
internal motions (Mac Low \& Norman 1993) and probably promote
turbulence in the swept-up gas. If they form stars or dense knots,
then these ballistic objects will emerge out the front of the
layer as it decelerates. They will not be bound to the layer by
gravitational forces because the deceleration is large and the
layer gravity is small at this stage (see Elmegreen 1989b; Nishi
1992; Nishi \& Kamaya 2000). This means that clusters and stars
embedded in swept-up shells or layers form by a slower mode of
instability.

Most triggered star formation in swept-up gas seems to occur after
the layer or shell has become significantly self-gravitating as a
whole, which means that the layer thickness is comparable to the
internal Jeans length, the Mach number is low, and the internal
density is only a factor of order 10 above the external density.
For example, strong supernova shocks rarely have young stars in
their compressed gas. They are too young for this and various
instabilities have fractured the gas into pieces smaller than the
Jeans mass (Chevalier \& Theys 1975; Vishniac 1983). Only the
oldest, largest, and slowest-moving shells have peripheral star
formation (e.g., Deul \& den Hartog 1990). This observation of
late collapse times is confirmed theoretically by direct
simulations of decelerating layers (Nishi \& Kamaya 2000), and by
a series of simulations with various environmental conditions that
show instability times in expanding shells always comparable to
about 0.2 times the external dynamical time,
$\left(G\rho\right)^{-1/2}$  (Elmegreen, Palou\v s, \& Ehlerov\'a
2002). When the collapse finally does begin, it tends to go very
quickly (Elmegreen \& Lada 1977; Nishi \& Kamaya 2000).

This timing result seems to provide an important link between
stellar compression as a trigger for star formation and the
duration-size relation. After large-scale gravitational
instabilities combined with supernova and other stellar pressures
generate turbulence and a corresponding hierarchical density
structure, the random stellar pressures continue to pound on the
gas and trigger stars.  The time scale for this triggering is
comparable to the average dynamical time in the region.
High-density regions, which tend to be small in a fractal gas,
have new stars triggered faster than low density regions, which
tend to be large. Thus, star formation moves around in a kpc-size
star complex for a long time, with one generation slowly
triggering others as a result of expanding HII regions, runaway
O-type stars and O stars that evaporated off of clusters, stray
supernova, and large-scale expansions into shells. At the same
time, star formation propagates around inside each smaller region
faster, forming OB subgroups inside of OB associations with the
same combination of HII region and supernova pressures, but now on
a shorter time scale because of the higher average GMC density.
More star formation may be triggered on even shorter times inside
the GMC cores. Because each region has a density-size relation
originally established by turbulent motions, any process of star
formation that operates on the local dynamical time, which
includes both spontaneous and triggered star formation,
contributes to a size-duration correlation for young stars that
mimics the size-crossing time relation in the gas.

An important difference arises in spiral density wave shocks,
where pre-existing clouds and gas crowd together to make a high
density of young stars (Nikola et al. 2001).  Here, there is
little deceleration so gravitational collapse occurs on the short
time scale of the compressed density rather than the long time
scale of the external density. The shock moves at a steady,
high-Mach speed through the galaxy, with greater Mach numbers and
greater compressions at smaller radii inside corotation. As a
result, ballistic condensations that form in the compressed gas
stay close to the shock front. Their separation from the front at
large radii, where the perpendicular component of the shock speed
is small (near corotation) and the compression is also small, is
comparable to their separation from the front at small radii,
where the perpendicular component of the speed is high and the
compression is high.  For this reason, HII regions and young stars
stay close to the spiral dust lane regardless of the distance from
corotation.

There are several numerical models now that include enough about
interstellar gas dynamics and star formation to reproduce most of
the observations, but no models yet reproduce all of the features
discussed here. Nomura \& Kamaya (2001) modelled triggered star
formation in a turbulent medium and got the size-duration
correlation on scales larger than $50$ pc as a result of cluster
drift at the initial cloud speed.  This is the characteristic
scale above which the turbulent speed exceeds the random walk
speed for their propagating star formation model. Chappell \&
Scalo (2001b) simulated star formation in collapsing and colliding
shells that were driven by other star formation. They also got
spatially correlated star fields and a size-duration relation
(Scalo \& Chappell 1999), but lacked the resolution to see
small-scale triggering at the edges of HII regions (see also the
review of models like this in V\'azquez-Semadeni et al. 2000).
Neither of these models discussed the $\Sigma_{crit}$ threshold
nor did they include spiral wave generation as a source of
structure and turbulence on larger scales. Wada \& Norman
(1999, 2001), Wada (2001), and Wada, Spaans, \& Kim (2000) modelled
a large fraction of a galaxy and reproduced the generation of
spiral arms and turbulent structures from gravitational
instabilities at the $\Sigma_{crit}$ threshold. They had triggered
star formation in turbulence-generated clouds, with much of the
triggering done by random or collective supernovae, as discussed
above.  They did not include stellar spiral waves, nor did they
check to see if there was a size-duration correlation or a Schmidt
law. Instead, they studied other statistical properties of their
results like the probability density function for gas density
(which was a log-normal) and variations in the star formation rate
over time.

\section{Towards a comprehensive theory of star formation}
\label{sect:theory}

There is no theory yet that can derive the star formation rate
from first principles.  If an expression like equation
(\ref{eq:schmidt}) is appropriate, then such a theory would
determine $\epsilon$ and $\omega$ as functions of scale and other
parameters. Most likely, simplifications like this are not
possible for a variable mixture of all star formation processes
(e.g., Chappell \& Scalo 2001b). Still, it is illustrative to
consider the Schmidt law again, this time from two other points
of view.

First consider the star formation rate in an individual cloud
core, where an equation like (\ref{eq:schmidt}) would be
\begin{equation}
\left({\rm SFR/Volume}\right)|_{core} \sim \epsilon_c\rho_c\omega_c .
\label{eq:schmidt2}\end{equation} Here, $\epsilon_c$ is the
efficiency of star formation inside each dense protostellar core,
$\rho_c$ is the average core density, and $\omega_c$ is the
dynamical rate inside the core, which we take to be
$\omega_c=\left(G\rho_c\right)^{1/2}$. The efficiency inside each
core is high enough (perhaps $\sim50\%$ -- Matzner \& McKee 2000)
that it does not have much influence on the overall star formation
rate; it is essentially constant. The dynamical rate $\omega_c$
does not have much influence either if we consider most star
formation becomes inevitable when the core density reaches a
certain value, like $\rho_c=10^5m(H_2)$ cm$^{-3}$ in the
solar neighborhood.  At this
density, big grains stop gyrating around the magnetic field
(Kamaya \& Nishi 2000), molecules begin to freeze onto grains
(Bergin et al. 2001), and the ionization fraction begins to drop
(Caselli et al. 2002). Also, the physical scale is so small that
most turbulent motions become subsonic, reducing any tendency for
turbulence to fragment the gas further (Goodman et al. 1998). Thus
the onset of star formation on a small scale can by marked
approximately by a certain density, $\rho_c$ (which may
vary with galactic environment), and this makes
$\omega_c$ well-defined, like $\epsilon_c$. In this sense, the
star formation rate, measured as a mass per unit volume per unit
time, is about constant in cores that form stars (assuming a
threshold density). The accretion rate onto a star is not constant
because it is the mass per unit time, without the volume, and so
depends on the core velocity dispersion which varies with scale
($M\omega\sim c^3/G$).

To extrapolate this core-theory to a galactic-scale theory, we
need to include all of the gas is that not forming stars. To do
this, we write
\begin{equation}
\left({\rm SFR/Volume}\right)|_{galaxy} 
=
\left({\rm SFR/Mass}\right)|_{core}\left(M_{core}/M_{galaxy}\right)
\rho = \epsilon_c\omega_c f_c\rho 
\label{eq:schmidt3}\end{equation} where
$\epsilon_c$, and $\omega_c$ have the same meaning as in
equation (\ref{eq:schmidt2}), but
now $f_c=M_{core}/M_{galaxy}$ 
accounts for the fraction of the gas on the large
scale that is in the form of dense, star-forming cores; $\rho$ is
the average density on this large scale, as in equation
(\ref{eq:schmidt}). If this scale is larger than a scale height,
then $\rho$ should be replaced by $\Sigma$ to get the star
formation rate per unit area; there should not be much change in
$\epsilon_c$ and $\omega_c$ with this substitution.  Now we have
$\epsilon_c\omega_cf_c$ as a replacement for $\epsilon\omega$ in
equation (\ref{eq:schmidt}); $\epsilon_c\omega_cf_c$ is
the product of
the star formation rate per unit mass inside unstable cores   
($\epsilon_c\omega_c$) and the mass fraction that
these cores represent ($f_c$).

One of the interesting aspects of writing the Schmidt law in this
way is that it emphasizes the geometry of the gas through the
fraction $f_c$. If this geometry a more universal property of the
ISM than the mixture of all the discrete physical processes that
enter into $\omega$ in the original theory, given by equation
(\ref{eq:schmidt}), then perhaps we have made some progress toward
finding a comprehensive theory.

There does appear to be a universal aspect to gas geometry.
Self-similar gas structures seem universal on scales between
star-forming cores and the galactic scale height, so we can cover
a wide range of galactic volume in our extrapolation from the
average density $\rho$ down to the star-forming density $\rho_c$.
What enters mathematically into this extrapolation is the
probability distribution function for density, $p(\rho)$. More
specifically, it is the probability distribution function for the
density of gas that is in the form of self-gravitating objects.
The pdf of density alone, which is equal to the probability that a
certain volume has a density $\rho$, has been discussed
extensively for turbulent and self-gravitating fluids
(V\'azquez-Semadeni 1994; Nordlund \& Padoan 1999; Klessen 2000;
Wada \& Norman 2001). It is remarkably invariant in the Wada
\& Norman (2001) simulation, having a
log-normal form even when the density structure comes from a
mixture of gravitational instabilities, star-formation pressures,
and turbulence.

For a log-normal, the fraction of all the mass with a density
larger than some threshold $\rho_c$ is given by the normalized
integral over $p(\rho)$,
\begin{equation}
f_c={{\int_{\rho_c}^{\infty} \rho p(\rho)
d\rho}\over{\int_{0}^{\infty} \rho p(\rho) d\rho}}
\label{eq:f0}\end{equation} where the normalized pdf is
\begin{equation}
p(\rho)={1\over{\left(2\pi\right)^{1/2}\Delta}}\exp\left(-0.5\left[
\ln\left(\rho/\rho_p\right) \right]^2/\Delta^2\right)
d\ln\rho/d\rho .
\end{equation} Here, $\rho_p$ is the density at the peak of the
log-normal, $\Delta$ is the Gaussian dispersion in the natural
log of the density, and
$d\ln\rho/d\rho=1/\rho$ 
converts the $\ln\rho$ interval
in the definition of the log-normal into a linear interval
for the integration. 

The fraction of this dense gas that is
self-gravitating may not be so easily determined, so the hard
problem of generalizing the Schmidt law remains. Nevertheless, 
equation (\ref{eq:f0}) for $f_c$ has several interesting
properties.

First,
$f_c$ approximately agrees with the observed value from the
Schmidt law in Kennicutt (1998b) if $\rho_c/<\rho>\sim10^{5}$ 
for average ISM density $<\rho>=\rho_p\exp\left(\Delta^2/2\right)$
using $\Delta=2.3$ from the
simulations in Wada \& Norman (2001). (Wada \& Norman's value of
1.41 for the dispersion uses an expression for $p$ that does not
have the 0.5 factor in the exponent, and which also has a 10-based
log in the exponent instead of a natural log.)
For example, with $\rho_c/<\rho>=10^3$, $10^4$, $10^5$,
and $10^6$ m$_{\rm H}$ cm$^{-3}$, 
the integral in equation 
(\ref{eq:f0})gives $f_c=10^{-1.5}$, $10^{-2.7}$,
$10^{-4.2}$, and $10^{-6.2}$, respectively.  

The observed value of $f_c$ may be obtained from Kennicutt's
(1998b) expression
\begin{eqnarray}
{\rm SFR}/{\rm Area}\sim 2.5\times10^{-4}
\left({{\Sigma}\over{{\rm M}_\odot \;{\rm pc}^{-2}}}\right)^{1.4}
\; {\rm M}_\odot {\rm kpc}^{-2}\;{\rm yr}^{-1} \sim
0.033\Sigma\Omega \label{eq:kenn}\end{eqnarray} for average star
formation rate $SFR$ out to the edge of the disk, average gas
surface density $\Sigma$ inside the edge, and rotation rate in the
outer disk $\Omega$.  The coefficient 0.033 comes from an observed
conversion rate of 21\% ($=2\pi\times0.033$) of the total gas mass
per orbit in the outer disk.  If this galaxy-average rate results
from a local Schmidt law of the same general form, ${\rm
SFR}(r)/{\rm Area}=\epsilon_{l}\Sigma(r)\Omega(r)$ for local
efficiency $\epsilon_{l}$, exponential disk
$\Sigma(r)=\Sigma_c{\rm exp}\left(-r/r_D\right)$, and flat
rotation curve, $\Omega(r)=\Omega_Dr_D/r$ then $\epsilon_l$ and
the factor 0.033 are related by an integral over SFR(r),
\begin{equation}
{{\int_0^{r_{edge}} SFR(r) 2\pi r dr}\over{\pi r_{edge}^2}}=0.033
{{\int_0^{r_{edge}}\Sigma(r) 2\pi rdr}\over{\pi r_{edge}^2}}
\Omega\left(r_{edge}\right) \end{equation} giving $\epsilon_l\sim
0.033r_D/r_{edge}$ for typical $r_D/r_{edge}\sim0.25$.

To convert this local expression with an angular rotation rate
into a local expression with a gas density, we use the fact that
most ISM densities are comparable to the tidal density, which
gives $\Omega\sim\left(2\pi G\rho/3\right)^{1/2}$.  This finally
gives the local Schmidt law, now written per unit volume with
average density $\rho$ instead of $\Sigma$,
\begin{equation}
{\rm SFR}/{\rm Volume} \sim
0.033\times0.25\times\left(2\pi/3\right)^{1/2}\rho
\left(G\rho\right)^{1/2}=0.012\rho\left(G\rho\right)^{1/2}.\end{equation}
Considering the definition of $f_c$ as the fraction of the
interstellar mass denser than $\rho_c$, we require
$0.012\left(G\rho\right)^{1/2}=\epsilon_cf_c\left(G\rho_c\right)^{1/2}$,
from which we derive
\begin{equation}f_c=0.012\epsilon_c^{-1}
\left(\rho/\rho_c\right)^{1/2}. \label{eq:f02}\end{equation}
Setting $\epsilon_c=0.5$ and $\rho_c/\rho=10^3$, $10^4$, $10^5$,
and $10^6$ as above, we get $f_c=10^{-3.1}$, $10^{-3.6}$,
$10^{-4.1}$, and $10^{-4.6}$, respectively.
Equations (\ref{eq:f0}) and (\ref{eq:f02}) give the same $f_c\sim10^{-4}$ at
$\rho_c/<\rho>\sim10^5$.  
This implies that the Schmidt law may result from star formation
in unstable cores with a density that is always about the same
factor times the average ISM density, namely, $10^5$.  

If instead of the Schmidt law,
with its areal rate dependence on the $\sim1.4$ power of 
column density, the total galactic star formation
rate is really proportional to 
the first power of the total gas mass, as suggested
by studies that find a constant efficiency
(cf. Sect. \ref{sect:schmidt}),
then we should use an observed rate given by
\begin{equation}
{\rm SFR}/{\rm Volume} \sim \rho/\tau\end{equation}
for constant gas consumption time $\tau\sim10^{9.6}$ years 
(Boselli, Lequeux, \& Gavazzi 2002).
In this case, $\epsilon_cf_c\left(G\rho_c\right)^{1/2}\sim1/\tau$,
giving 
\begin{equation}
f_c\sim\left[\epsilon_c\tau\left(G\rho_c\right)^{1/2}\right]^{-1},
\end{equation}
which has values of 
$f_c=10^{-3.0}$, $10^{-3.5}$,
$10^{-4.0}$, and $10^{-4.5}$
for
$\epsilon_c=0.5$ and 
$\rho_c=10^3$, $10^4$, $10^5$,
and $10^6$ m(H$_2$) cm$^{-3}$, respectively.
Again the theoretical and observational values of
$f_c$ agree for a fraction $\sim10^{-4}$ of the total
gas mass in dense cores, but now 
the threshold core density $\rho_c$
has a constant value of $\sim10^5$ m(H$_2$) cm$^{-3}$
instead of a relative value of $\rho_c/<\rho>\sim10^5$. 

A second implication of equation (\ref{eq:f0}) 
is that most of the gas does not evolve
monotonically toward collapse and the formation of stars.  If all of the
relevant processes have dynamical timescales, then the rate at which
parcels of gas move around in the pdf is $\propto\rho^{1/2} p(\rho)$. This
function (or a similar one made with a time scale proportional to another
power of density) is also a log-normal in density (write $\rho^{1/2}$
as $\exp\left(0.5\ln\rho\right)$ and then complete the square in the
exponent), with no density-independent 
part except near the peak. This means that
each parcel of gas has to have some probability of moving both up and
down in density. If all of the mass with a density $\rho_1$ evolves at
the dynamical rate toward a higher density $\rho_2$, and then continues
to evolve at the dynamical rate to an even higher density $\rho_3$,
then $\rho_1^{1/2}p(\rho_1)$ would have to equal $\rho_2^{1/2}p(\rho_2)$
in order to conserve mass. However, $\rho^{1/2}p(\rho)$ decreases as a
log-normal for higher densities. Thus
there can be no monotonic progression at the dynamical rate
toward higher densities. The
fact that $\rho^{1/2}p(\rho)$ is higher at lower densities (to the
right of the peak) means that the probability for both compression and
rarefaction has to be higher at lower densities too.  Most clumps in a
medium with this pdf will therefore get either compressed or dispersed
after a dynamical time.  The progression toward higher densities is a
random walk along the abscissa of the $\rho^{1/2}p(\rho)$ curve.  Once a
sufficiently high $\rho$ is reached that the collapse becomes monotonic,
the pdf should change to a power law like $p(\rho)\propto\rho^{-0.5}$.

When star formation is viewed in this way, we can see more clearly
the importance of random events. The probability that a clump is
destroyed by an external flow goes down with increasing density,
in proportion to $\rho^{1/2}p(\rho)$, but it never goes to zero
until this pdf is violated, which presumably happens when local
gravity becomes so strong that collapse begins. Thus a small clump
in one region might survive the random bursts of pressures from
external flows around it while an identical clump in another
region might not. Observation of the clump alone, without any
attention to the distorting flows in the environment, will not
give the whole picture of its future evolution.

This probabilistic nature of star formation is the reason why
Schmidt-type laws are deceptively simple.  They suggest that
certain processes are deterministic, when in fact they are not --
given our limited knowledge of all the flows and forces in the
region.  Stochasticity is an unavoidable implication of the
presence of interstellar turbulence in star-forming gas.  The
initial stellar mass function may be stochastic in this sense: we
cannot write an equation for the mass of a star that forms in gas
with certain bulk properties, but can only give the probability
distribution function for all possible stellar masses
(e.g., Elmegreen 1999).  

Stochasticity in star formation may have another interesting
effect.  It may lead to time variations that have a fractal
quality, with both large- and small-scale, and short- and
long-time excursions.  Time-average rates are not 
well-defined when temporal variations are fractal. The financial
stock market is an example of a stochastic system with a fractal
time behavior (Mandelbrot 1997).  Time variations in the star
formation rates in galaxies are well known (Rocha-Pinto et al.
2000), with the largest variations appearing in the smallest
systems (Hunter 1997), as expected for a stochastic process. This
does not identify any particular mechanism for star formation,
such as stochastic self-propagating star formation (Gerola,
Seiden, \& Schulman 1980), but it may implicate the general role
of turbulence in establishing the geometric properties of the gas,
including the density pdf and the intermittent nature of the
flows.  The time variations are not likely to be large in the
sense that a quiescent galaxy turns into a starburst because of
intermittency in turbulence (starbursting dwarfs may be
interacting with external matter anyway -- Pustilnik et al. 2001),
but they can be mild and continuous, as are the variations in real
dwarf galaxies (van Zee 2001), having as a signature only the lack
of specific and identifiable causes for each variation.

The previous discussion expresses the small-scale view of star
formation, first emphasizing what we know about dense clumps and
then extrapolating to the average ISM using universal geometric
properties. Such an approach may be useful as an illustration of
star formation efficiency and turbulent stochastic effects, but it
is probably not useful as a means of deriving the global star formation
rate from first principles. The bottleneck in the conversion of
ambient gas into stars is at the largest scale, which dominates
the overall dynamical time because of its low density. Nearly any
physical process can form stars on a small scale, considering the
very short times involved, without greatly affecting the overall
star formation rate on the large scale.

The opposite point of view may be more productive. Considering the
sensitivity of the star formation rate to $\Sigma/\Sigma_{crit}$,
the long-term and large-scale average rate should equal the rate
at which the column density increases above the threshold value.
This increase comes from vertical infall and in-plane accretion
driven by viscous and spiral arm torques. Gravitational
instabilities sensitive to $\Sigma/\Sigma_{crit}$ generate spiral
arms and turbulence, and these arms generate torques which lead to
accretion.  The accretion increases $\Sigma$ but does not change
$\Sigma_{crit}$ much, so in a steady state, all of the excess
column density above $\Sigma_{crit}$ should form stars, regardless
of the mechanisms involved.  The link between the star formation
rate and the accretion rate can also lead to exponential disk
structure (Lin \& Pringle 1987).

The problem with this approach is that there is yet no
comprehensive theory of interstellar torques and accretion rates
that can be used to get $d\Sigma/dt$. The advantage of it, if there
were such an accretion theory, is that it can readily account for
the large influence that external perturbing galaxies have on the
star formation rate through transient enhancements in spiral arm
torques. Indeed, the major episodes of star formation in the Milky
Way disk correspond to those in the LMC, and both may have
occurred at the times of our closest approaches (Rocha-Pinto et
al. 2000). It seems futile to try to formulate such variations in
terms of $f_c$ and a generalized Schmidt law, but not at all
unreasonable in terms of a simple rule like ${\rm SFR}/{\rm
Area}\sim d\left(\Sigma-\Sigma_{crit}\right)/dt$.

\section{Conclusions}

The empirical laws of star formation on a large scale suggest that
clouds form by gravitational instabilities and compression from
turbulence and stellar pressures when an equilibrium cool phase of
interstellar matter is available.  Gravitational instabilities
make spiral arms and beads of star formation along the arms, and
they pump energy into turbulence. The instabilities also lead
directly to star formation by monotonic collapse in some cases,
particularly when the ambient density is high or the
environment is quiet. Turbulence structures the gas over a
wide range of scales, making a pervasive multifractal network.
Turbulence compression can lead directly to star formation too, in
the converging parts of the flow.  This may be an important
formation mechanism for individual stars in the dense cores of
molecular clouds. The onset of star formation in the whole cores,
however, appears to be related more strongly to the presence of
external pressures from other stars. These pressures are inferred
from the cometary or flattened morphologies of gas structures 
enclosing embedded young clusters.  The empirical laws and
correlations apparently work on large and intermediate scales,
even when star formation is triggered on small scales, because
the triggering time is usually comparable to the local dynamical time.

This interpretation explains the origin of the $\Sigma_{min}$ and
$\Sigma_{crit}$ thresholds for galaxy-wide star formation, the
$\Sigma^{1.4}$ power-law dependence for the average star formation
rate, the self-similar structure of interstellar gas and young stars
and the associated size-duration correlation for young star fields,
and the common appearance of young clusters at the tips of filamentary
clouds or at the edges of HII regions and other pressure sources.
The emphasis on spiral instabilities as a significant source of turbulence
also helps explain why $\Sigma\sim\Sigma_{crit}$ in most galaxy disks:
spiral instabilities are more directly related to the ratio of these
quantities than is the star formation rate.  

The Schmidt law suggests that star formation can be viewed in either
of two ways: (1) as a process on the scale of individual star-forming
cores with a high efficiency and a high rate per unit mass, giving an
overall galactic rate equal to this high rate multiplied by the core
mass fraction, or (2) as a process on the scale of the galactic disk
with an inefficient conversion of gas into stars at the large-scale
dynamical rate for the ISM.  Both viewpoints give the same result when
the density distribution function is a log-normal like that found by Wada
\& Norman (2001).  When the Schmidt law is satisfied, star formation
is saturated to its maximum possible value given the low fraction of
the gas ($10^{-4}$) that is allowed to be dense in a turbulent medium.
Detailed triggering mechanisms for individual clusters do not seem
to matter for the overall star formation rate.  Whatever mixture of
physical processes is involved, the fastest rate that stars can form
is given by the total gas mass in a dense state divided by the collapse
time at that state.  

Considerations of star formation with a threshold galactic column density
suggest that the rate should have a long-time average
that is equal to the accretion rate above this threshold.  Models of
accretion rates are currently not detailed enough to make this point of
view useful.

Acknowledgements: This work was supported by NSF grant
AST-9870112.

\end{document}